\documentclass{birkjour}
\usepackage[noadjust]{cite}
\usepackage{xcolor}
\RequirePackage[all]{xy}

\newtheorem{thm}{Theorem}[section]

\newtheorem{prop}[thm]{Proposition}
\theoremstyle{definition}
\newtheorem{defn}[thm]{Definition}
\theoremstyle{remark}
\newtheorem{rem}[thm]{Remark}
\newtheorem{comment}[thm]{Comment}

\numberwithin{equation}{section}

\newcommand{\bn}{\mathbf{n}}
\newcommand{\be}{\mathbf{e}}
\newcommand{\bs}{\mathbf{s}}

\newcommand{\bZ}{\mathbf{Z}}

\newcommand{\BZ}{\mathbb{Z}}
\newcommand{\BR}{\mathbb{R}}

\newcommand{\BK}{\mathbb{K}}
\newcommand{\BI}{\mathbb{I}}

\newcommand{\ed}{\end{document}}
\begin{document}

%-------------------------------------------------------------------------
% editorial commands: to be inserted by the editorial office
%
%\firstpage{1} \volume{228} \Copyrightyear{2004} \DOI{003-0001}
%\seriesextraline{This is the Concrete Title of this Book\br H.E. R and S.T.C. W, Eds.}
%\firstpage{1}

\title[Approximation, orthogonal polynomials, and Toda equations]
 {Hermite--Pad\'{e} approximation, \\multiple orthogonal polynomials,\\ and multidimensional Toda equations}
%----------Author 1
\author[Adam Doliwa]{Adam Doliwa}
%\author[Birkh\"auser \textit{et al.}]%
%{Birkh\"{a}user Publishing Ltd.}
%
\address{%	
Faculty of Mathematics and Computer Science\\
University of Warmia and Mazury\\
ul. S\l oneczna 54, 10-710 Olsztyn\\ Poland}
\email{doliwa@matman.uwm.edu.pl}
%
%\thanks{No thanks}

\date{\today}
%\dedicatory{Last Revised:\\ \today}
%%% ----------------------------------------------------------------------
\begin{abstract}
We review recent results on the connection between Hermite--Pad\'e approximation problem, multiple orthogonal poly\-no\-mials, and multidimensional Toda equations in continuous and discrete time. In order to motivate interest in the subject we first present a pedagogical introduction to the classical, by now, relation between the Pad\'e approximation problem, orthogonal polynomials, and the Toda lattice equations. We describe also briefly generalization of the connection to interpolation problems and to the non-commutative algebra level.
\end{abstract}

%\label{page:firstblob}
%%% ----------------------------------------------------------------------
\maketitle
%%% ----------------------------------------------------------------------
%\tableofcontents

\section{Introduction}
The Toda lattice equations~\cite{Toda-TL} form a paradigmatic example of a completely integrable system~\cite{AblSi,ZMNP} whose role goes much beyond~\cite{Flaschka,Kac-vMoerbeke,Moerbeke,Symes,Olshanetsky-Perelomov,Moser} its initial mechanical context. It allows for a generalization to a partial differential equation~\cite{Mikhailov} whose origin is deeply rooted in classical differential geo\-me\-try of surfaces and their transformations~\cite{Darboux-TS}. From the other side, its diffe\-ren\-ce version \cite{Hirota} showed that certain  numerical algorithms~\cite{Baker,Gragg,Brezinski-PTA-OP,CuytWuytack,BultheelvanBarel}, studied by distinguished  mathematicians of XIXth century ~\cite{Brezinski,Cauchy,Jacobi,Frobenius,Hermite}, have a lot in common~\cite{GRP,Hirota-1993,NagaiTokihiroSatsuma,PGR-LMP} with discrete integrable systems~\cite{IDS}. 

Moreover, integrable difference version, by Hirota~\cite{Hirota-2dT}, of the gene\-ra\-lized Toda system~\cite{Mikhailov}, contains as symmetry reductions or appropriate limits many of known integrable equations, notably the whole  Kadomtsev--Petviashvili hierarchy~\cite{DKJM}. See also \cite{Miwa,Shiota,KNS-rev} for other aspects of the Hirota system, its non-commutative version~\cite{Nimmo-NCKP}, geometric interpretation~\cite{DCN,Dol-Des} and symmetry structure~\cite{Dol-AN}. 

The goal of the paper is to review recent results on relation~\cite{Doliwa-Siemaszko-2,Doliwa-NHP} of the Hirota system to the Hermite--Pad\'{e} approximation problem~\cite{Hermite-P,Baker,Paszkowski,DD-DC-1}. We present also, as closely related~\cite{AptekarevDerevyaginMikiVanAssche} subject,  the theory of multiple orthogonal polynomials~\cite{Aptekarev,Ismail,VanAsche-nn-mop,VanAssche}. Our main motivation is the connection of orthogonal polynomials theory to numerical algorithms, in particular its structural affinity with the Pad\'{e} approximants, and relation to the theory of integrable systems, see also \cite{ManoTsuda,Nagao-Yamada} for discussion of other aspects of the relation.

It is difficult to overestimate the role of orthogonal polynomials~\cite{Chihara} in theoretical physics and applied mathematics. They can be encountered when solving, by separation of variables, partial differential equations of the classical field theory or the quantum mechanics~\cite{Nikiforov-Suslov-Uvarov}.
Another source are problems of probability theory leading~\cite{Schoutens,Iglesia} to the difference equations of hypergeometric type or various $q$--analogs of the above ones \cite{Ismail}. 
In mathematical physics they can be found in the spectral theory of operators in Hilbert space~\cite{Akhiezer,Szego,Geronimus}. More recently they were associated with the theory of representations of groups and algebras (including quantum groups) \cite{Vilenkin-Klimyk,Klimyk-Schmudgen}. They have found application in the theory of random matrices~\cite{Deift} and are used to construct special solutions to Painlev\'{e} equations~\cite{AdlervanMoerbeke,VanAssche}. 

The structure of the paper is as follows. In Section~\ref{sec:Pade-OP} we present basic elements of the Pad\'{e} approximation theory together with its connection to the theory of orthogonal polynomials and Toda equations in continuous and discrete time. We take also the opportunity to introduce basic aspects of analytical theory of integrable equations. The next Section is devoted to Hermite--Pad\'{e} approximation and related difference equations~\cite{Paszkowski,DD-DC-1,DD-DC-2} which turn out to provide integrable reduction of the Hirota system, and  appeared almost at the same time as~\cite{Hirota-2dT}. In the last Section we present analogous results of the theory of multiple orthogonal polynomials, and we close the paper with remarks on generalization of the above subjects to the non-commutative level~\cite{Doliwa-NHP} and to interpolation problem~\cite{Doliwa-NAMTS}.

\section{Essentials of Pad\'{e} approximants and orthogonal polynomials} \label{sec:Pade-OP}
Below we present the well known~\cite{IDS} relation between the Pad\'e approximation problem, orthogonal polynomials, and the Toda lattice equations. We will pay attention to these elements of the theory which motivate both the spectral aspects of the theory of integrable systems and existence of infinite family of commuting symmetries.
\subsection{The Pad\'{e} problem and Frobenius identities as the discrete-time Toda lattice equations}
Given power series
$f(x) = \sum_{i=0}^\infty f_i x^i$ with $f_0 \neq 0$, and given pair of non-negative integers $(m,n)\in \mathbb{N}_0^2$,  find polynomials $P_{m,n}(x)$ and $Q_{m,n}(x)$ with $\deg P_{m,n} \leq m$, $\deg Q_{m,n} \leq n$ such that their ratio  approximates $f(x)$ to the highest order possible 
\begin{equation} 
f(x) = \frac{P_{m,n}(x)}{Q_{m,n}(x)} + O(x^{m+n+1}).
\end{equation}

\begin{prop}
The answer to the Pad\'{e} problem can be obtained if we rewrite the above equation in the form 	
\begin{equation} 
Q_{m,n}(x) \left( \sum_{i=0}^\infty f_i x^i \right) - P_{m,n}(x) \equiv 0 \quad \mod \quad x^{m+n+1},
\end{equation}	
and collect coefficients at different powers of $x$ to get a system of \emph{linear} homogeneous equations. This leads to the Jacobi formulas~\cite{Jacobi,Gragg}
%\begin{small}
\begin{gather}
	P_{m,n}(x)  = \left| \begin{matrix}
	F_m(x) & x F_{m-1}(x) & \cdots & x^n F_{m-n}(x) \\
	f_{m+1} & f_m    & \cdots & f_{m-n+1}  \\
	\vdots       &  \vdots & \ddots & \vdots \\
	f_{m+n} & f_{m+n-1} &\cdots & f_m
	\end{matrix} \right| , \\ \qquad \label{eq:Q-PA}
	Q_{m,n}(x)   = \left| \begin{matrix}
	1    & x & \cdots & x^n \\
	f_{m+1} & f_m    & \cdots & f_{m-n+1}  \\
	\vdots       &  \vdots & \ddots & \vdots \\
	f_{m+n} & f_{m+n-1} &\cdots & f_m
	\end{matrix}
	\right| ,
\end{gather}
%\end{small}
where $F_k(x) = \sum_{i=0}^k f_i x^i$ are polynomial approximants of the series, and  by convention $F_k(x) \equiv 0 $ for $k<0$.
\end{prop}

It is well known that calculating determinants from the very definition is not efficient numerically. The standard procedure is based on the Jacobi theorem, known also as the Sylvester identity \cite{Hirota-book}. In application to the 
square matrix $M$, and when by $M_{[i;j]}$ we denote the matrix with $i$-th row and $j$-th column removed, it reads
\begin{equation*}
\det M \cdot \det M_{[i_1 i_2; j_1 j_2]} = \det M_{[i_1;j_1]} \cdot \det M_{[i_2;j_2]} - \det M_{[i_1;j_2]} \cdot \det M_{[i_2;j_1]},
\end{equation*}
where also we take $i_1 < i_2$, and $ j_1 < j_2$.

The following relations, derived by Frobenius~\cite{Frobenius}, which involve an arbitrary linear combination 
$ W_{m,n}(x) = 
a(x)P_{m,n}(x)  + b(x) Q_{m,n}(x)$ 
of the polynomials in question,
and the determinants of the Hankel type built off the series coefficients
%\begin{small}
\begin{equation} \label{eq:D-PA}
\Delta_{m,n} = \left| \begin{matrix}
f_{m} & f_{m-1}    & \cdots & f_{m-n+1}  \\
f_{m+1} & f_{m} &\cdots & f_{m-n+2} \\
\vdots       &  \vdots & \ddots & \vdots \\
f_{m+n-1} & f_{m+n-2} &,\cdots & f_m
\end{matrix} \right| ,
\end{equation}
%\end{small}
can be obtained~\cite{Baker,Gragg} with the help of the Sylvester identity

%\begin{center}
%	\framebox{%
%		\begin{minipage}{0.95\linewidth}
	\begin{small}
	\begin{align} \label{eq:F-1}
	\Delta_{m+1,n} W_{m,n}(x) - \Delta_{m,n} W_{m+1,n}(x) & = x \Delta_{m+1,n+1} W_{m,n-1}(x), \\ 
	\label{eq:F-2}
	\Delta_{m,n} W_{m,n+1}(x) - \Delta_{m,n+1} W_{m,n}(x) & = x \Delta_{m+1,n+1} W_{m-1,n}(x), \\ 
%	\tag{L1}
	\label{eq:L1}
	\Delta_{m+1,n} W_{m,n+1}(x) - \Delta_{m,n+1} W_{m+1,n}(x) & = x \Delta_{m+1,n+1} W_{m,n}(x), \\ %\tag{L2}
	\label{eq:L2}
	\Delta_{m+1,n} W_{m-1,n}(x) + \Delta_{m,n+1} W_{m,n-1}(x) & =  \Delta_{m,n} W_{m,n}(x), \\
%	\tag{DTTL}
	\label{eq:DTTL}
	\Delta_{m+1,n} \Delta_{m-1,n} + \Delta_{m,n+1} \Delta_{m,n-1} & =  \Delta_{m,n}^2, \\
	\label{eq:F-6}
	W_{m+1,n}(x) W_{m-1,n}(x) + W_{m,n+1}(x) W_{m,n-1}(x) & =   W_{m,n}^2(x).
	\end{align} 
	\end{small}
%	\end{minipage}}
%\end{center}
The nonlinear equation~\eqref{eq:DTTL} is known, in the theory of discrete integrable systems, as the discrete-time Toda lattice equation~\cite{Hirota-2dT,Toda-TL}. It can be obtained, independently of the numerical context (in particular allowing the discrete variables $m,n$ to take arbitrary integer values) by elimination of the function $W_{m,n}(x)$ from the system \eqref{eq:L1}-\eqref{eq:L2}. 

\subsection{Orthogonal polynomials and discrete-time Toda lattice}
Let $\mu$ be a positive measure on the real line, consider the corresponding system of monic polynomials subject to orthogonality relations
\begin{equation} \label{eq:orthogonality}
\int_\mathbb{R} Q_r(x) Q_s(x) d\mu(x) = 0, \qquad r\neq s, \qquad Q_r(x) = x^r + \dots , 
\end{equation}
The Cauchy--Stieltjes transform of the measure and the moments, for which we assume that all they exist, are defined by
\begin{equation} \label{eq:CS}
S(z) =  \int_\mathbb{R} \frac{d\mu(x)}{z-x} = \sum_{k=0}^\infty \frac{\nu_k}{z^{k+1}}, \qquad \nu_k = \int_{\mathbb{R}} x^k d\mu(x). 
\end{equation}
\begin{comment}

If the measure is decomposed into absolutely continuous and discrete parts  
\begin{equation}
d\mu(x) = u(x)dx + d\mu_s(x),
\end{equation}
then
\begin{enumerate}
\item the weight is given by 
\begin{equation}
u(x) = \lim_{\varepsilon\to 0^+}\frac{1}{\pi} \mathrm{Im}S(x+i\varepsilon);
\end{equation}
\item the singular part $\mu_s$ is concentrated on the set \begin{equation}
\{ x_0 \colon \lim_{x\to x_0} \mathrm{Im} S(x) = \infty \};
\end{equation}
\item the mass of any point is given by 
\begin{equation}
\mu(\{ x_0 \}) = \lim_{\varepsilon\to 0^+} \varepsilon \mathrm{Im} S(x_0 + i\varepsilon ).
\end{equation}
\end{enumerate}
\end{comment}
Conditions 
\begin{equation} \label{eq:orth-pol}
\int x^k Q_s(x)d\mu(x) = 0, \qquad k=0, \dots , s-1,
\end{equation}
interpreted as linear equations 
for coefficients of the polynomial lead to 
\begin{equation}  \label{eq:Q-OP}
Q_s(x) = \frac{1}{D_s} \left| \begin{matrix}
	\nu_0 & \nu_1 & \nu_2 & \dots & \nu_s \\
	\nu_1 & \nu_2 & \nu_3 & \dots & \nu_{s+1} \\
	\vdots & \vdots & \ddots  & & \vdots \\
	\nu_{s-1} & \nu_s & \nu_{s+1} & \dots & \nu_{2s-1} \\
	1 & x & x^2 & \dots & x^s
	\end{matrix} \right| , 
\end{equation}
where $D_s$ is the algebraic complement of the element $x^s$ in the right-lower corner
\begin{equation} \label{eq:D-OP}
	D_s = \left| \begin{matrix}
	\nu_0 & \nu_1 &  \dots & \nu_{s-1} \\
	\nu_1 & \nu_2 &  \dots & \nu_{s} \\
	\vdots & \vdots & \ddots  & \vdots \\
	\nu_{s-1} & \nu_s &  \dots & \nu_{2s-2} 
	\end{matrix} \right| ;
\end{equation}
notice the strong similarity of \eqref{eq:Q-OP} and \eqref{eq:D-OP} to \eqref{eq:Q-PA} and \eqref{eq:D-PA}, respectively. The correspondence goes further, in particular one can show~\cite{Ismail} that the polynomial $P_{s-1}(x)$ of degree $s-1$, defined by
	\begin{equation} \label{eq:P-Q}
P_{s-1}(x) = \int \frac{Q_s(z) - Q_s(x)}{z-x} d\mu(x),
\end{equation}
satisfies the following condition
	\begin{equation*}
	S(x) Q_{s}(x) - P_{s-1}(x) = O(x^{-s-1}), \qquad x \to \infty ,
	\end{equation*}
which can be derived (see \cite{VanAsche-HPAO} for details) from Pad\'{e} approximation at infinity of the Cauchy--Stieltjes transform \eqref{eq:CS}.

To obtain the discrete-time Toda lattice equations let us consider the following variation of the measure $d\mu_t(x) = x^t d\mu(x)$, $t\in \mathbb{N}_0$. Then the corresponding simple evolution of the moments $\nu_{s,t} = \nu_{s+t}$ gives
\begin{equation*}
D_{s,t} = \left| \begin{matrix}
\nu_t & \nu_{t+1} &  \dots & \nu_{t+ s-1} \\
\nu_{t+1} & \nu_{t+2} &  \dots & \nu_{t+s} \\
\vdots & \vdots & \ddots  & \vdots \\
\nu_{t+s-1} & \nu_{t+s} &  \dots & \nu_{t+2s-2} 
\end{matrix} \right| 
\end{equation*} 
and implies, by comparison with equation~\eqref{eq:D-PA}, that the above determinants satisfy the discrete-time Toda lattice equations \eqref{eq:DTTL} with variables $n=s$, $m=t+s-1$
\begin{equation} \label{eq:DTTL-D}
D_{s-1,t+1} D_{s+1,t-1} = D_{s,t-1} D_{s,t+1} - D_{s,t}^2.
\end{equation} 

Let us discuss the connection in more detail~\cite{AptekarevDerevyaginMikiVanAssche} within the orthogonal polynomials approach, which will be generalized in Section~\ref{sec:multiple-OP}. Notice that the difference of two monic polynomials $Q_{s+1,t}(x)$ and $xQ_{s,t+1}$ is a polynomial of degree $s$ orthogonal to $x^k$, $k=0,1, \dots , s-1$, with respect to the measure $d\mu_t(x)$, and therefore must be proportional to $Q_{s,t}$
\begin{equation} \label{eq:lin-Q-A}
xQ_{s,t+1}(x) = Q_{s+1,t}(x) + A_{s,t}Q_{s,t}(x).
\end{equation}
The coefficient of proportionality 
\begin{equation} \label{eq:A-D}
A_{s,t} = \frac{D_{s+1,t+1}D_{s,t}}{D_{s+1,t}{D_{s,t+1}}}
\end{equation}
can be found by multiplying both sides of equation~\eqref{eq:lin-Q-A} by $x^s$, integrating with respect to $d\mu_t(x)$ and noticing that formula~\eqref{eq:Q-OP} implies that
\begin{equation}
\int x^s Q_{s,t} d\mu_t(x) = \frac{D_{s+1,t}}{D_{s,t}}.
\end{equation}
\begin{rem}
	Equations \eqref{eq:lin-Q-A}-\eqref{eq:A-D} can also be obtained from the Sylvester identity applied to the matrix of $Q_{s+1,t}$ with respect to its first and last columns, and the two last rows.
\end{rem}
Similar reasoning applied to the difference $Q_{s,t}$ and $Q_{s,t+1}$ gives equations
\begin{equation} \label{eq:lin-Q-B}
Q_{s,t}(x) = Q_{s,t+1}(x) + B_{s,t}Q_{s-1,t+1}(x),
\end{equation}
where
\begin{equation} \label{eq:B-D}
B_{s,t} = \frac{D_{s+1,t}D_{s-1,t+1}}{D_{s,t}{D_{s,t+1}}}.
\end{equation}
By substituting \eqref{eq:lin-Q-A} to \eqref{eq:lin-Q-B} we obtain the standard three-term formula for orthogonal polynomials in the form
\begin{equation} \label{eq:3-term-AB}
xQ_{s,t}(x) = Q_{s+1,t}(x) + (A_{s,t} + B_{s,t}) Q_{s,t}(x) + A_{s-1,t}B_{s,t}Q_{s-1,t}(x),
\end{equation}
and by substituting \eqref{eq:lin-Q-B} to \eqref{eq:lin-Q-A} we get
\begin{equation*}
xQ_{s,t+1}(x) = Q_{s+1,t+1}(x) + (A_{s,t} + B_{s+1,t}) Q_{s,-t+1}(x) + A_{s,t}B_{s,t}Q_{s-1,t+1}(x).
\end{equation*}
Comparison of both equations gives another form of discrete-time Toda lattice equations
\begin{equation} \label{eq:DTTL-AB}
A_{s,t+1} + B_{s,t+1} = A_{s,t} + B_{s+1,t}, \qquad 
A_{s-1,t+1} B_{s,t+1} = A_{s,t} B_{s,t} . 
\end{equation}
\begin{rem}
Notice that the second part of the equations is directly obtained from the parametrizations~\eqref{eq:A-D} and~\eqref{eq:B-D}, while the first part is a further consequence of the bilinear form  \eqref{eq:DTTL-D} of the discrete-time Toda lattice equations.
\end{rem}

\subsection{The continuous-time Toda lattice equations}	
Assume that measure undergoes evolution~\cite{Kac-vMoerbeke,Moser} of the form
\begin{equation} \label{eq:mu-t}
d\mu(x,t) = e^{-xt} d\mu(x), \qquad t \in \BR_+,
\end{equation}
well known in the theory of continuous-time birth and death processes~\cite{LedermannReuter,KarlinMcGregor}.
The corresponding evolution of the moments and of the determinants reads
\begin{equation} \label{eq:nu-D-evol}
\dot{\nu}_s(t)= -\nu_{s+1}(t) \qquad \Rightarrow \quad \dot{D}_s(t) = - \tilde{D}_s(t), 
\end{equation}
where the dot means derivative with respect to the time variable $t$, and
\begin{equation} \label{eq:tD-OP}
\tilde{D}_s = \left| \begin{matrix}
\nu_0 & \nu_1 &  \dots & \nu_{s-2} & \nu_{s} \\
\nu_1 & \nu_2 &  \dots & \nu_{s-1} & \nu_{s+1} \\
\vdots & \vdots & \ddots  & & \vdots \\
\nu_{s-2} & \nu_{s-1} & &  \nu_{2s-4} & \nu_{2s-2} \\
\nu_{s-1} & \nu_s &  \dots & \nu_{2s-3} & \nu_{2s-1} 
\end{matrix} \right| , 
\end{equation}
is the determinant of the matrix of $D_{s+1}$ with the last row and penultimate column removed.	
\begin{prop}
The Sylvester identity and elementary properties of determinants imply the Toda lattice equations in the Hirota form
\begin{equation} \label{eq:CTTL}
\ddot{D}_s(t) D_s(t) = \dot{D}_s(t)^2 + D_{s+1}(t) D_{s-1}(t).
\end{equation}
\end{prop}
In order to obtain another version of the continuous-time Toda lattice equations \eqref{eq:CTTL} it is convenient to consider first the celebrated three-term recurrence relation for orthogonal polynomials~\cite{Akhiezer,Szego,Ismail}, which in a different form appeared already~\eqref{eq:3-term-AB}. The orthogonality  conditions \eqref{eq:orthogonality} give
\begin{equation} \label{eq:3-term}
xQ_s(x) = Q_{s+1}(x) + b_s Q_s(x) + a_s Q_{s-1}(x),
\end{equation}
where	
	\begin{equation*}
	a_s = \frac{\int [Q_s(x)]^2 d\mu(x)}{\int [Q_{s-1}(x)]^2 d\mu(x)}, \qquad b_s = 
	\frac{\int x [Q_s(x)]^2 d\mu(x)}{\int [Q_{s}(x)]^2 d\mu(x)} .
	\end{equation*}
Representation \eqref{eq:Q-OP} of the polynomials gives
\begin{equation} \label{eq:a-OP}
\int x^s Q_s(x)d\mu(x) = \frac{D_{s+1}}{D_s} \qquad \Rightarrow \quad a_s = \frac{D_{s+1} D_{s-1}}{D_s^2}, 
\end{equation}
and 
\begin{equation} \label{eq:b-OP}
Q_s(x) = x^s - \frac{\tilde{D}_s}{D_s} x^{s-1} + \dots
\qquad \Rightarrow \quad b_s = \frac{\tilde{D}_{s+1}}{D_{s+1}} - \frac{\tilde{D}_{s}}{D_{s}} ,
\end{equation}
which together with \eqref{eq:CTTL} implies that the coefficients of the three-term relation satisfy 
	\begin{align} \label{eq:TF-1}
	\dot{a}_s(t) & =  a_s(t)(b_{s-1}(t) - b_s(t)), \\
	\label{eq:TF-2}
	\dot{b}_s(t) & =  a_s(t) - a_{s+1}(t),
	\end{align}
	which are the Toda lattice equations in the form given by Flaschka~\cite{Flaschka}. 
	
Equivalently, one can observe~\cite{Kac-vMoerbeke,Moser} that the functions $Q_s(x,t)$ satisfy equations
\begin{equation} \label{eq:evol-Q}
\frac{\partial Q_s(x,t)}{\partial t} = a_s(t) Q_{s-1}(x,t). 
\end{equation}
Then the Toda lattice equations \eqref{eq:TF-1}-\eqref{eq:TF-2} arise from compatibility of the recurrence~\eqref{eq:3-term} with the evolution \eqref{eq:evol-Q}.

\begin{rem}
	After the change of variables 
	\begin{equation}
	a_s(t) = e^{q_{s-1}(t)-q_s(t)},
	\end{equation} 
	which allows for identification $b_s(t) = \dot{q}_s(t)$ we obtain the original form of Toda lattice equations~\cite{Toda-TL}
	\begin{equation} \label{eq:Toda-Q}
	\ddot{q}_s(t) = e^{q_{s-1}(t)-q_s(t)} -e^{q_{s}(t)-q_{s+1}(t)}. 
	\end{equation}
	In application to the theory of orthogonal polynomials we put $q_{-1}(t) \equiv - \infty$, what gives $a_0(t)\equiv 0$.	
\end{rem}		

\begin{comment}
The idea that simple evolution of the measure reflects in non-linear evolution of the recurrence coefficients is crucial in the inverse spectral (or scattering) technique of solving integrable equations \cite{GGKM,AblSi,ZMNP}.
\end{comment}
Define the following matrices, the so called Lax pair,
%\begin{small}
\begin{equation*}
L(t) = \begin{pmatrix}
b_0(t) & 1 & 0 &  \hdots  \\
a_1(t) & b_1(t) & 1 &  \hdots  \\
0 & a_2(t) & b_2(t) &     \\
\vdots & \vdots&  &  \ddots
\end{pmatrix}, \qquad
A(t) = \begin{pmatrix}
0 & 0 & 0&   \hdots  \\
a_1(t) & 0 & 0&  \hdots  \\
0 & a_2(t) & 0 &     \\
\vdots & \vdots&   & \ddots 
\end{pmatrix},
\end{equation*}
%\end{small}
such that equations \eqref{eq:3-term}, \eqref{eq:evol-Q} can be rewritten in terms of the vector-function $Q(x,t)=(Q_0(x,t),Q_1(x,t), Q_2(x,t), \dots )^T$ as
\begin{equation} \label{eq:Lax-TL}
L(t) Q(x,t) = x Q(x,t), \qquad \frac{\partial Q(x,t)}{\partial t} = A(t) Q(x,t)
\end{equation}
The first equation has the form of the eigen-value problem, while the second one is adjusted such that the support of the spectral measure is invariant with respect to the time evolution. The compatibility condition of \eqref{eq:Lax-TL}
\begin{equation}
\dot{L}(t) + [ L(t), A(t)] = 0,
\end{equation}
is equivalent to equations~\eqref{eq:TF-1}-\eqref{eq:TF-2}, and is called their Lax form.
\begin{rem}
The half-infinite chain of the Toda lattice can be considered as a reduction of the bi-infinite chain. Other basic reductions are provided by restrictions to finite or periodic chains.  The structure of their Lax pair is closely related to Lie algebra of type $\mathfrak{sl}$ and its affine version, correspondingly. Similar versions of the Toda lattice equations can be constructed for any simple Lie algebra or its affine version~\cite{Olshanetsky-Perelomov}.
\end{rem}
	
\begin{comment} 
Define sequence of times $(t_1,t_2,\dots)$ by the following evolutions of the measure  
\begin{equation}
d\mu(x,t_k) = e^{-x^k t_k} d\mu(x), \qquad k=1,2,3, \dots ,
\end{equation} 
which commute with evolution with respect to the original time $t=t_1$ and with each other. The corresponding nonlinear differential equations for the recursion coefficients are called the Toda lattice hierarchy. Existence of infinite system of mutually commuting symmetries of a given integrable partial differential equation is another fundamental concept~\cite{DKJM,Olver}  of soliton theory.
\end{comment}

\section{Hermite--Pad\'{e} approximation and integrability}
Below we present a generalization of the Pad\'{e} approximation problem together with the corresponding generalizations of the Frobenius identities.
\subsection{Formulation and solution of the Hermite--Pad\'{e} problem}
The interest of Hermite in algebraic approximation, which was related to his famous proof~\cite{Hermite,Hermite-P} of transcendence of the Euler number $e$, led him to consider the following problem, which we present in the contemporary formulation \cite{DD-DC-1}. We remark that Charles Hermite was doctoral supervisor of Henri Pad\'{e}.
\begin{defn}
Given $m$ elements $(f_1(x),\dots , f_m(x))$ of the algebra $\BK[[x]]$ of formal series in variable $x$ with coefficients in the field $\BK$, 
\begin{equation}
f_i(x) = \sum_{j=0}^\infty f^i_j x^j, 
\end{equation}
and given multi-index $\bn=(n_1,\dots , n_m)\in\BZ_{\geq -1}^m$, where we also write $|\bn|= n_1 + \dots + n_m$. A solution of the corresponding \emph{Hermite--Pad\'{e} problem} is every system of polynomials $(Q_{1,\bn}(x), \dots , Q_{m,\bn}(x))$ in $\BK[x]$, not all equal to zero, with corresponding degrees $\deg Q_{i,\bn}(x) \leq n_i$, $i=1,\dots , m$ (degree of the zero polynomial by definition equals $-1$), and such that 
\begin{equation}
Q_{1,\bn}(x)f_1(x) + \dots + Q_{m,\bn}(x) f_m(x) \equiv 0 \quad  \mod \quad x^{|\bn|+m-1}.
\end{equation}
\end{defn}
Solution of the Hermite--Pad\'{e} problem is given as follows \cite{DD-DC-1}. Consider the matrix $\mathcal{M}_\bn$ of $r = (|\bn|+m-1)$ rows and $r+1 = (|\bn|+m)$ columns 
%\begin{small}
\begin{equation}
\mathcal{M}_\bn = \begin{pmatrix}
f^1_0 &     & 0 & \! & \!& f^m_0 &   & 0 \\
f^1_1 & \ddots    &   & \! \cdots & \! \cdots & f^m_1 & \ddots  &   \\
\vdots &   &  f^1_0 & \! & \! & \vdots &    &  f^m_0 \\
\vdots &   &  \vdots & \! \cdots &\! \cdots & \vdots &  & \vdots\\
f^1_{r-1} & \cdots & f^1_{r - n_1 -1} & \! & \! & f^m_{r-1} & \cdots & f^m_{r - n_m -1}
\end{pmatrix},
\end{equation}
%\end{small}
which consists of $m$ (possibly empty) blocks, the $i$th block is composed out of $n_i +1$ columns depending on the coefficients of $f_i(x)$ only.
By supplementing $\mathcal{M}_\bn$ at the bottom by the line
\begin{equation*}
\left( f_1(x), xf_1(x), \dots , x^{n_1} f_1(x), \cdots , \cdots , f_m(x), x f_m(x), \dots , x^{n_m} f_m(x) \right),
\end{equation*}
and calculating the determinant of the resulting square matrix in two ways we obtain the following identity
\begin{equation} \label{eq:Z-f-Delta}
Z_{1,\bn}(x) f_1(x) + \dots + Z_{m,\bn}(x) f_m(x) = x^{|\bn|+m-1} \sum_{j=0}^\infty \Delta^{(j-1)}_\bn x^j, 
\end{equation}
where each $Z_{k,\bn}(x)\in \BK[x]$, $k=1,\dots m $, is a polynomial of degree not exceeding $n_k$. It is given explicitly by the determinant
\begin{equation} \label{eq:Z-det}
Z_{k,\bn}(x) =  \left|
\begin{smallmatrix}
f^1_0 &     & 0 & \! & & f^k_0 & & 0 && \!& f^m_0 &   & 0 \\
f^1_1 & \ddots    &   & \! \cdots & \cdots & f^k_1  &\ddots &  &\cdots &\cdots  \!& f^m_1 & \ddots  &   \\
\vdots &   &  f^1_0 & \!& & \vdots && f^k_0 & & \!& \vdots &     &  f^m_0 \\
\vdots &   &  \vdots & \!\cdots &\cdots &&&&\cdots &\cdots  \! & \vdots &  & \vdots\\
f^1_{r-1} & \cdots & f^1_{r - n_1 -1} &\! & &f^k_{r-1}& \cdots & f^k_{r-n_k - 1}& &  \!&  f^m_{r-1} & \cdots & f^m_{r - n_m -1} \\
0 & \cdots & 0 & \! \cdots & \cdots &1 & \cdots & x^{n_k} & \cdots &\cdots  \!& 0 & \cdots & 0
\end{smallmatrix} \! \right|,
\end{equation}
of the matrix $\mathcal{M}_\bn$ supplemented at the bottom by the line 
\begin{equation*}
%\label{eq:Xk}
(0,\dots, 0, \dots \; \dots , 1, x , \dots , x^{n_k}, \dots \; \dots , 0, \dots ,0),
\end{equation*}
consisting of zeros except for the $k$th block of the form $1, x, \dots , x^{n_k}$. Each
$\Delta^{(j)}_\bn$ is the determinant of the matrix $\mathcal{M}_\bn$ supplemented by the line
\begin{equation*} % \label{eq:bN}
%b_{r+j+1} = 
\left( f^1_{r+j+1}, \dots , f^1_{r+j+1-n_1}, \cdots , \cdots , f^m_{r+j+1},  \dots , f^m_{r+j+1-n_m} \right),
\end{equation*} 
as the last row, respectively. In particular, the first coefficient of the sum on the right hand side of equation \eqref{eq:Z-f-Delta} equals 
\begin{equation}
\Delta_\bn = \Delta^{(-1)}_\bn = \left| \begin{matrix}
f^1_0 &     & 0 & & & f^m_0 &   & 0 \\
f^1_1 & \ddots    &   & \cdots & \cdots & f^m_1 & \ddots  &   \\
\vdots &   &  f^1_0 & & & \vdots &    &  f^m_0 \\
\vdots &   &  \vdots & \cdots &\cdots & \vdots &  & \vdots\\
f^1_{r-1} & \cdots &   & & &  & \cdots & f^m_{r - n_m -1} \\
f^1_{r} & \cdots & f^1_{r - n_1} & & & f^m_{r} & \cdots & f^m_{r - n_m }
\end{matrix} \right| .
\end{equation}

The above system of polynomials $ \bZ_\bn(x) = (Z_{1,\bn}(x), \dots , Z_{m,\bn}(x))$ is called  \emph{the canonical Hermite--Pad\'{e} form of degree~$\bn$}. When the polynomials $Z_{k,\bn}(x)$ are of the maximal possible order then the corresponding solution of the Hermite--Pad\'{e} problem is unique, up to an overall factor, and such multi-index $\bn$ is called \emph{normal}.
Because the leading term of the polynomial $Z_{k,\bn}(x)$ reads
\begin{equation} \label{eq:leading-term-Z}
Z_{k,\bn}(x) = (-1)^{(n_{k+1} + \dots + n_m) - m+k}\Delta_{\bn-\be_k} x^{n_k} + \dots,
\end{equation}
therefore this happens if the determinants $\Delta_{\bn-\be_k}$, $k=1,\dots,m$, do not va\-nish. If all multi-indices $\bn\in\BZ^m_{\geq -1}$ are normal then such a system of series $(f_1(x), \dots , f_m(x))$ is called \emph{perfect} \cite{Mahler-P}, what we assume in the sequel.

\subsection{The Paszkowski reduction of the Hirota system}
Application of Sylvester's identity to the determinants $Z_{k,\bn+\be_i+\be_j}(x)$, $i\neq j$, $k = 1,\dots ,m$,  with two bottom rows and the last columns of the $i$th and $j$th blocks, gives \cite{DD-DC-2}
\begin{equation} \label{eq:Z-Delta}
\bZ_{\bn+\be_i +\be_j}(x) \Delta_\bn = \bZ_{\bn +\be_j}(x) \Delta_{\bn+\be_i} - 
\bZ_{\bn+\be_i}(x) \Delta_{\bn+\be_j}, \qquad i<j .
\end{equation}
The above equation is known, in a broader context of integrable systems, as the linear problem~\cite{DJM-II} to the discrete Kadomtsev--Petviashvili system in Hirota's form~\cite{Hirota}, which reads
\begin{equation} \label{eq:DD-Hirota}
\Delta_{\bn+\be_i + \be_j} \Delta_{\bn+\be_k} -
\Delta_{\bn+\be_i + \be_k} \Delta_{\bn+\be_j} +
\Delta_{\bn+\be_j + \be_k} \Delta_{\bn+\be_i} = 0, 
\end{equation}
where we assume $i < j < k$. The above system can be obtained, in the integrable equations approach, by elimination
of the wave function $\bZ_\bn(x)$ from the equations \eqref{eq:Z-Delta}. In the context of Hermite--Pad\'{e} equations it can be derived using Sylvester's identity.  The same equation is satisfied~\cite{Doliwa-Siemaszko-2} by each polynomial $Z_{\ell,\bn}(x)$, $\ell = 1,\dots ,m$
\begin{equation} 
\label{eq:ZZ-Hirota}
\begin{split}
Z_{\ell, \bn+\be_i + \be_j}(x) & Z_{\ell, \bn+\be_k}(x) -
Z_{\ell,\bn+\be_i + \be_k}(x) Z_{\ell,\bn+\be_j}(x) +\\
& Z_{\ell,\bn+\be_j + \be_k}(x) Z_{\ell,\bn+\be_i}(x) = 0, \qquad i < j < k . 
\end{split}
\end{equation}
\begin{comment}
Equation of type \eqref{eq:Z-Delta} appears as \eqref{eq:L2} in the set of standard Frobenius identities \eqref{eq:F-1}-\eqref{eq:F-6}. Two other equations \eqref{eq:DD-Hirota} and \eqref{eq:ZZ-Hirota} involve three discrete variables, and do not have analogs in the set.  
\end{comment}

In \cite{Paszkowski} it was shown by Paszkowski, using analyticity and uniqueness properties of solutions of the Hermite--Pad\'{e} problem (see ~\cite{Doliwa-Siemaszko-2} for determinantal proof), that the canonical Hermite--Pad\'{e} form satisfies the following equation
\begin{equation}
\label{eq:Paszkowski-Frobenius}
x {\bZ_\bn(x)} \Delta_\bn = \bZ_{\bn + \be_1}(x) \Delta_{\bn - \be_1} + \dots + \bZ_{\bn + \be_m}(x) \Delta_{\bn - \be_m}
\end{equation}
the corresponding generalization of a Frobenius identity~\eqref{eq:L1}. Consequences of \eqref{eq:Z-Delta} and \eqref{eq:Paszkowski-Frobenius} are \cite{Paszkowski,Doliwa-Siemaszko-2} two other analogs of Frobenius identities \eqref{eq:DTTL} and \eqref{eq:F-6}
\begin{gather}
\label{eq:Paszkowski-Frobenius-DD}
\Delta_\bn^2 = \Delta_{\bn + \be_1} \Delta_{\bn - \be_1} + \dots + \Delta_{\bn + \be_m} \Delta_{\bn - \be_m},\\
Z_{\ell,\bn}(x)^2 = Z_{\ell,\bn + \be_1}(x)  Z_{\ell,\bn - \be_1}(x) + \dots +  Z_{\ell,\bn + \be_m}(x)  Z_{\ell,\bn - \be_m}(x).
\end{gather}

The Paszkowski constraint \eqref{eq:Paszkowski-Frobenius-DD} provides integrable reduction~\cite{Doliwa-Siemaszko-2} of the Hirota system \eqref{eq:DD-Hirota} in the more general context of soliton theory, where the discrete variables are arbitrary integers. The full system of equations generalizes the discrete-time Toda lattice equations to arbitrary number of variables. In Section~\ref{sec:multiple-OP} we present the interpretation of corresponding analogs of the equations within the context of multiple orthogonal polynomials~\cite{AptekarevDerevyaginMikiVanAssche}.

\subsection{Simultaneous Pad\'{e} approximants}
Another generalization of the Pad\'{e} approximants for several series is provided by the simultaneous approximation problem~\cite{Baker}, also considered by Hermite~\cite{Hermite,Hermite-P}. Given 
$m$ series $(f_1(x),\dots , f_m(x))$ in $\BK[[x]]$
and $m$ integers $\bn=(n_1,\dots , n_m)\in \BZ_{\geq -1}^m$. A solution of the corresponding \emph{simultaneous Pad\'{e} problem} is every system of polynomials $(P_{1,\bn}(x), \dots , P_{m,\bn}(x))$ in $\BK[x]$, not all equal to zero, with corresponding degrees $\deg P_{i,\bn}(x) \leq |\bn| - n_i$, $i=1,\dots , m$, and such that for all pairs $i\neq j$
\begin{equation}
P_{i,\bn}(x) f_j(x) - P_{j,\bn}(x) f_i(x) \equiv 0 \quad \mod \quad x^{|\bn| + 1}. 
\end{equation}
Also here a solution in terms of determinants is known~\cite{BeckermannLabahn}, however we will not present its details because there exists a remarkable connection, found by Mahler~\cite{Mahler-P}, between solutions of the Hermite--Pad\'{e} approximation problem and the simultaneous Pad\'{e} problem for the same system of series.

\begin{thm} \label{th:Mahler}
	Define $m\times m$ matrix $\mathcal{Q}_\bn(x)$ whose $i$th row consists of the solution of Hermite--Pad\'{e} problem with degrees $\bn - \be + \be_i$ 
	\begin{equation}
	[\mathcal{Q}_\bn(x)]_{ij} = Q_{j,\bn - \be + \be_i}(x),
	\end{equation}
	where $\be = \be_1 + \be_2 + \dots + \be_m$,
	normalized such that diagonal elements are monic polynomials. Similarly, define $m\times m$ matrix $\mathcal{P}_\bn(x)$ whose $i$th row consists of the solution of simultaneous Pad\'{e} problem with degrees $\bn - \be_i$ 
	\begin{equation}
	[\mathcal{P}_\bn(x)]_{ij} = P_{j,\bn - \be_i}(x),
	\end{equation}
	normalized such that diagonal elements are monic polynomials, then
	\begin{gather}
	\mathcal{Q}_\bn(x) \mathcal{P}_{\bn}(x)^{T} = x^{|\bn|} \BI, \\ 
	|\mathcal{Q}_\bn(x)| = x^{|\bn|}, \qquad |\mathcal{P}_\bn(x)| = x^{(m-1)|\bn|}.
	\end{gather}
\end{thm}
The above duality implies that both the Hirota system~\eqref{eq:DD-Hirota} and the Paszkowski constraint~\eqref{eq:Paszkowski-Frobenius-DD} should be present within the theory of the simultaneous Pad\'{e} approximation problem. The corresponding detailed analysis will be presented elsewhere.

\section{Multiple orthogonal polynomials and multidimensional Toda lattice}
\label{sec:multiple-OP} 
In this Section we consider a generalization~\cite{Aptekarev,NikishinSorokin} of the theory of orthogonal polynomials, which has found recently application in mathematical physics and theory of random processes~\cite{AptekarevDerevyaginMikiVanAssche,BleherKuijlaars,Kuijlaars,VanAsche-HPAO,VanAsche-nn-mop}. Following~\cite{AptekarevDerevyaginMikiVanAssche} we will show its connection to the simultaneous Pad\'{e} approximation problem and relation to multidimensional generalizations of the Toda systems.

\subsection{Multiple orthogonal polynomials} 
\begin{defn}
Given $r$ positive measures $\mu^{(j)}$, $j=1,\dots , r$, on the real line, and given $r$ non-negative integers $\bs = (s_1,\dots , s_r) \in\mathbb{N}_0^r$. The \emph{multiple orthogonal polynomial of the index $\bs$} with respect to the measures is the polynomial $Q_{\bs}(x)$ (usually normalized to monic) of degree $|\bs|=s_1 + \dots + s_r$
which satisfies the following orthogonality conditions with respect to the measures
\begin{equation} \label{eq:orth-r}
\int Q_{\bs}(x) x^k d\mu^{(j)}(x) = 0, \qquad k=0,1,\dots , s_j - 1 ,
\end{equation}
generalizing that of \eqref{eq:orth-pol}.
\end{defn}
The above conditions give $|\bs|$ linear equations for $|\bs|$ coefficients of the polynomial.
The Cauchy--Stieltjes transforms of the measures and the corresponding moments, for which we assume that all exist, are defined by
\begin{equation} 
S^{(j)}(z) = \int_\mathbb{R} \frac{d\mu^{(j)}(x)}{z-x} =
 \sum_{k=0}^\infty \frac{\nu_{k}^{(j)}}{z^{k+1}}  , \qquad \nu_{k}^{(j)} = \int_{\mathbb{R}} x^k d\mu^{(j)}(x).
\end{equation}
The polynomial $Q_\bs(x)$ exists and is unique provided the determinant
\begin{equation} \label{eq:D-r}
\qquad \quad
D_{\bs}  = \left| \begin{matrix}
\nu_{0}^{(1)} & \nu_{1}^{(1)} &  \dots & \nu_{|\bs|-1}^{(1)}\\
\vdots & \vdots & \ddots   & \vdots \\
\nu_{s_1 - 1}^{(1)} & \nu_{s_1}^{(1)} &  \dots & \nu_{ |\bs|+ s_1 - 2}^{(1)} \\
\cdots & \cdots &   & \cdots \\
\nu_{0}^{(r)} & \nu_{1}^{(r)} &  \dots & \nu_{|\bs|-1}^{(r)}\\
\vdots & \vdots & \ddots   & \vdots \\
\nu_{s_r - 1}^{(r)} & \nu_{s_r}^{(r)} &  \dots & \nu_{ |\bs| + s_r - 2}^{(r)} 
\end{matrix} \right| ,
\end{equation}
does not vanish; then the corresponding multi-index $\bs$ is called normal. In such case we have
\begin{equation} \label{eq:Q-r}
Q_{\bs}(x) = \frac{1}{D_{\bs}} \left| \begin{matrix}
\nu_{0}^{(1)} & \nu_{1}^{(1)}  &  \dots & \nu_{|\bs|}^{(1)} \\
\vdots & \vdots & \ddots   & \vdots \\
\nu_{s_1 - 1}^{(1)}  & \nu_{s_1}^{(1)}  &  \dots & \nu_{ |\bs|+ s_1 - 1}^{(1)}  \\
\cdots & \cdots &   & \cdots \\
\nu_{0}^{(r)} & \nu_{1}^{(r)} &  \dots & \nu_{|\bs|}^{(r)}\\
\vdots & \vdots & \ddots   & \vdots \\
\nu_{s_r - 1}^{(r)} & \nu_{s_r}^{(r)} &  \dots & \nu_{ |\bs| + s_r - 1}^{(r)} \\
1 & x & \dots & x^{|\bs|}
\end{matrix} \right| .
\end{equation}
\begin{rem}
	The matrices in \eqref{eq:D-r} or \eqref{eq:Q-r} are composed out of $r$ row-blocks, each $j$th block comes from equations~\eqref{eq:orth-r} for the $j$th measure.
\end{rem}

Define $r$ polynomials $P_{\bs}^{(j)} (x)$, $j=1,\dots ,r$,  by analogs of equation~\eqref{eq:P-Q}
\begin{equation*}
P_{\bs}^{(j)} (x) = \int \frac{Q_{\bs}(z) - Q_{\bs}(x)}{z-x} d\mu^{(j)} (x),
\end{equation*}
which, due to orthogonality relations \eqref{eq:orth-r}, satisfy the simultaneous approximation conditions at infinity, see \cite{Aptekarev} for details,
\begin{equation*}
S^{(j)} (x) Q_{\bs}(x) - P_{\bs}^{(j)} (x) = O(x^{-s_j-1}), \qquad x\to \infty.
\end{equation*}

\begin{rem}
	There exists another system of polynomials satisfying certain orthogonality condition with respect to the  measures $d\mu^{(j)}(x)$, $j=1,\dots , r$, which leads to the Hermite--Pad\'{e} approximation problem, see \cite{Aptekarev,NikishinSorokin} for details. 
\end{rem}

\subsection{Multidimensional Toda lattice system in discrete time}
Let us assume that the measures evolve~\cite{AptekarevDerevyaginMikiVanAssche} according to the equation 
\begin{equation}
d\mu_{t}^{(j)}(x) = x^t d\mu^{(j)}(x), \qquad t\in \mathbb{N}_0.
\end{equation} 
By the reasoning analogous to that in derivation of equations~\eqref{eq:lin-Q-A} and \eqref{eq:lin-Q-B}  we find 
\begin{align} \label{eq:lin-Q-A-r}
xQ_{\bs,t+1}(x) & = Q_{\bs + \be_j,t}(x) + A^{(j)}_{\bs,t} Q_{\bs,t}(x), \qquad j=1,\dots r,\\ \label{eq:lin-Q-B-r}
Q_{\bs,t}(x) & = Q_{\bs,t+1}(x) + \sum_{j=1}^r B^{(j)}_{\bs,t} Q_{\bs - \be_j,t+1}(x),
\end{align}
where
\begin{equation} \label{eq:AB-Q-r}
A^{(j)}_{\bs,t} = \frac{D_{\bs + \be_j,t+1}D_{\bs,t}}{D_{\bs+\be_j , t} D_{\bs ,t+1}}, \qquad
B^{(j)}_{\bs,t} = \frac{D_{\bs + \be_j,t}D_{\bs - \be_j,t+1}}{D_{\bs, t} D_{\bs ,t+1}},
\end{equation}
and 
\begin{equation}
D_{\bs,t} = \left| \begin{matrix}
\nu_{t}^{(1)}  & \nu_{t+1}^{(1)}  &  \dots & \nu_{t+|s|-1}^{(1)}  \\
 \vdots   &  \vdots     &        & \vdots          \\
\nu_{t+s_1 - 1}^{(1)}  & \nu_{t+s_1}^{(1)}  &  \dots & \nu_{t+|s|+s_1 -2}^{(1)}  \\
 \hdots   &  \hdots    &         &  \hdots  \\
\nu_{t}^{(r)}  & \nu_{t+1}^{(r)}  &  \dots & \nu_{t+|s|-1}^{(r)}  \\
\vdots & \vdots & \ddots  & \vdots \\
\nu_{t+s_{r}-1}^{(r)}  & \nu_{t+s_r}^{(r)}  &  \dots & \nu_{t+|s|+s_r-2}^{(r)}  
\end{matrix} \right| ,
\end{equation}
and the polynomial $Q_{\bs,t}(x)$ is given by the corresponding modification of formula \eqref{eq:Q-r}.
\begin{prop}
The compatibility of the system \eqref{eq:lin-Q-A-r} for $j\neq k$ gives the following relations between its coefficients
\begin{equation} \label{eq:AA-r}
A_{\bs,t+1}^{(j)} - A_{\bs,t+1}^{(k)} = A_{\bs+\be_k,t}^{(j)} - A_{\bs+\be_j,t}^{(k)}, 
\quad A_{\bs+\be_k,t}^{(j)} A_{\bs,t}^{(k)} = A_{\bs+\be_j,t}^{(k)} A_{\bs,t}^{(j)}.
\end{equation}
\end{prop}
\begin{rem}
	Notice that the second part of the above equations is directly obtained from the parametrization \eqref{eq:AB-Q-r}, while the first part is a further consequence of the bilinear Hirota's relations (compare with \eqref{eq:DD-Hirota})
	\begin{equation} \label{eq:D-jk-t}
	D_{\bs+\be_j + \be_k,t} 	D_{\bs,t+1} = 	D_{\bs+ \be_k,t+1} 	D_{\bs+\be_j,t } - 	D_{\bs+\be_j, t+1} 	D_{\bs+\be_k, t}, \qquad j<k, 
	\end{equation}
	which can be derived using Sylvester's identity.
\end{rem}
Three copies of equations \eqref{eq:D-jk-t} built for pairs of indices from $i<j<k$ imply the bilinear equations~\eqref{eq:DD-Hirota} for the lattice space variables
\begin{equation} \label{eq:D-ijk}
D_{\bs+\be_i + \be_j,t} 	D_{\bs+\be_k,t} -	D_{\bs+ \be_i + \be_k,t} 	D_{\bs+\be_j,t } + 	D_{\bs+\be_j + \be_k, t} 	D_{\bs+\be_i, t} = 0. 
\end{equation}
From the other hand, equations \eqref{eq:lin-Q-A-r} imply 
\begin{equation}
Q_{\bs+\be_j,t}(x) - Q_{\bs+\be_k,t} = \left( A^{(k)}_{\bs,t} - A^{(j)}_{\bs,t}\right) Q_{\bs,t}(x),
\end{equation}
which due to \eqref{eq:D-jk-t} can be brought to the form
\begin{equation}
Q_{\bs+\be_j,t}(x) - Q_{\bs+\be_k,t} = \frac{D_{\bs+\be_j + \be+k,t} D_{\bs,t}}{D_{\bs+\be_j,t}D_{\bs + \be+k,t}} Q_{\bs,t}(x), \qquad j<k,
\end{equation}
which is the linear problem for \eqref{eq:D-ijk}.
	The symmetry between variables $s_i$, $i=1,\dots , r$, and the discrete-time variable $t$ can be restored by introducing new variable $s_{0}$ instead of $t$ 
	\begin{equation} \label{eq:s-t}
	s_{0} = t+s_1 + \dots + s_r.  
	\end{equation}
\begin{prop}
The compatibility of \eqref{eq:lin-Q-A-r} and \eqref{eq:lin-Q-B-r} gives, in addition to \eqref{eq:AA-r}, equations
\begin{align} \label{eq:A+B-r}
A_{\bs,t+1}^{(j)} + \sum_{k=1}^r B_{\bs,t+1}^{(k)} & = A_{\bs,t}^{(j)} + \sum_{k=1}^r B_{\bs+\be_j,t}^{(k)}, 
\\ \label{eq:AB-r}
A_{\bs-\be_j,t+1}^{(j)} B_{\bs,t+1}^{(j)} & = A_{\bs,t}^{(j)} B_{\bs,t}^{(j)},\\ \label{eq:BAA-r}
B^{(k)}_{\bs,t}\left( A^{(k)}_{\bs,t} - A^{(j)}_{\bs,t} \right) & = B^{(k)}_{\bs + \be_j,t}\left( A^{(k)}_{\bs-\be_k,t+1} - A^{(j)}_{\bs-\be_k,t+1} \right).
\end{align}
\end{prop}
\begin{rem}
	Equations~\eqref{eq:AB-r} can be directly obtained from the parametrization \eqref{eq:AB-Q-r}, while equations \eqref{eq:BAA-r} result as a further consequence of the bilinear Hirota system~\eqref{eq:D-jk-t}. Finally, equations~\eqref{eq:A+B-r} follow  from the bilinear equation stated in~\cite{AptekarevDerevyaginMikiVanAssche}
	\begin{equation} \label{eq:MDTE-D}
	D_{\bs,t}^2 = 	D_{\bs,t+1} 	D_{\bs,t-1 } - \sum_{k=1}^r 	D_{\bs+\be_k, t-1} 	D_{\bs-\be_k, t+1},
	\end{equation}
	which is~\cite{Doliwa-Siemaszko-2} the Paszkowski constraint~\eqref{eq:Paszkowski-Frobenius-DD} in the new variables \eqref{eq:s-t}.
\end{rem}

\subsection{Multidimensional Toda lattice system in continuous time}
For multiply orthogonal polynomials we have following analogs~\cite{VanAsche-nn-mop} of the three term relation~\eqref{eq:3-term}
\begin{equation} \label{eq:3-term-r}
x Q_\bs (x) = Q_{\bs + \be_j} + b_{\bs}^{(j)} Q_{\bs}(x) + \sum_{k=1}^r a_{\bs}^{(k)} Q_{\bs - \be_k}(x), \qquad j=1,2,\dots , r,
\end{equation}
where the recurrence coefficients are given by formulas
\begin{equation} \label{eq:a-b-r}
a_{\bs}^{(j)} =  \frac{D_{\bs + \be_j} D_{\bs - \be_j}}{D_{\bs}^2},\qquad 
b_{\bs}^{(j)} = \frac{\tilde{D}_{\bs + \be_j}}{D_{\bs + \be_j}} - \frac{\tilde{D}_{\bs}}{D_{\bs}} .
\end{equation}
Here $\tilde{D}_\bs$ is determinant of the matrix in \eqref{eq:Q-r} with the last row and penultimate column removed. Equivalently, it is the determinant $D_{\bs+\be_j}$ with the last row of $j$th block and the penultimate column deleted. In the above we used the expansion 
\begin{equation} \label{eq:Q-exp-r}
Q_\bs(x,t) = x^{|\bs|} - \frac{\tilde{D}_\bs(t)}{D_\bs(t) } x^{|\bs|-1} + \dots,
\end{equation}
\begin{prop}
In contrary to the single measure case $r=1$ the coefficients cannot be given arbitrarily but should satisfy~\cite{VanAsche-nn-mop} the compatibility conditions of \eqref{eq:3-term-r}
\begin{align} \label{eq:mo-1}
\quad \; b_{\bs+\be_k}^{(j)} - b_{\bs + \be_j}^{(k)} & = b_{\bs}^{(j)} - b_{\bs}^{(k)},\\  \label{eq:mo-2}
b_{\bs}^{(k)} b_{\bs + \be_k}^{(j)} - b_{\bs}^{(j)} b_{\bs + \be_j}^{(k)}  & = \sum_{i=1}^r \left( a_{\bs+\be_k}^{(i)} - a_{\bs+\be_j}^{(i)}\right),\\  \label{eq:mo-3}
a_{\bs + \be_k}^{(j)}( b_{\bs -\be_j}^{(j)} - b_{\bs - \be_j}^{(k)}) & = a_{\bs}^{(j)}(b_{\bs}^{(j)} - b_{\bs}^{(k)}).
\end{align}
\end{prop}

Assume that the measures evolve in continuous time 
\begin{equation}
	\label{eq:mu-t-r}
	d\mu^{(j)}(t,x) = e^{-xt} d\mu^{(j)}(x), \qquad t \in \BR_+, \qquad j=1,\dots ,r,
\end{equation}
like in equation~\eqref{eq:mu-t}. Then induced evolution of the moments reads
\begin{equation} \label{eq:evol-nu-r}
\dot{\nu}_{k}^{(j)}(t) = - \nu_{k+1}^{(j)}(t),
\end{equation}
what implies that
\begin{equation} \label{eq:D-t-r}
\dot{D}_{\bs}(t) = - \tilde{D}_\bs(t) ,
\end{equation}	
what leads to
\begin{equation}
b_{\bs}^{(j)}(t) = \frac{\dot{D}_\bs(t)}{D_\bs(t)} - 
\frac{\dot{D}_{\bs+\be_j}(t)}{D_{\bs+\be_j}(t)} .
%= \frac{d}{dt} \left( \log %\frac{D_{\bs}(t)}{D_{\bs+\be_j}(t)} \right).
\end{equation}
Notice that for $r\geq 2$ we have also for $i<j$
\begin{equation} \label{eq:dot-D-r}
D_{\bs + \be_i + \be_j}(t) D_{\bs}(t) = \dot{D}_{\bs+\be_i}(t) D_{\bs+\be_j}(t) - \dot{D}_{\bs+\be_j}(t) D_{\bs+\be_i}(t),
\end{equation}
which is direct consequence of \eqref{eq:D-t-r} and of the Sylvester identity applied to $D_{\bs + \be_i + \be_j}(t)$, its two last rows of the $i$th and $j$th blocks, and two last columns. For $r\geq 3$ three such equations for pairs from $i<j<k$ lead to the Hirota equation~\eqref{eq:D-ijk}.

Derivative of the expression~\eqref{eq:Q-r} with the moments evolving according to \eqref{eq:evol-nu-r} leads~\cite{AptekarevDerevyaginMikiVanAssche}, with the help of the Pl\"{u}cker determinantal identities, to 
\begin{equation} \label{eq:d-Q-r}
\frac{\partial Q_{\bs}(x,t)}{\partial t} =  \sum_{j=1}^r a_{\bs}^{(j)}(t) Q_{\bs - \be_j}(x,t) .
\end{equation}
Collecting in equation \eqref{eq:d-Q-r} coefficients at $x^{|\bs|-1}$ and using the expansion \eqref{eq:Q-exp-r}
together with equation~\eqref{eq:D-t-r}, one can obtain~\cite{AptekarevDerevyaginMikiVanAssche} the multidimensional analog of the continuous-time Toda equations
\begin{equation} \label{eq:D-mT}
\ddot{D}_\bs(t) D_\bs(t) - \left[ \dot{D}_\bs(t) \right]^2 = \sum_{j=1}^r D_{\bs+\be_j}(t) D_{\bs-\be_j}(t).
\end{equation}
\begin{prop}
The compatibility conditions between the discrete system~\eqref{eq:3-term-r} and the evolution equation \eqref{eq:d-Q-r} read~\cite{AptekarevDerevyaginMikiVanAssche} 
\begin{align} \label{eq:dot-a-r}
\dot{a}_{\bs}^{(j)}(t) & = a_{\bs}^{(j)}(t) \left( b_{\bs - \be_j}^{(j)}(t) - b_{\bs}^{(j)}(t)  \right),\\
\label{eq:dot-b-r}
\dot{b}_{\bs}^{(j)}(t) & = \sum_{k=1}^r \left(a_{\bs}^{(k)}(t) - a_{\bs + \be_j}^{(k)}(t) \right),
\end{align}
where also the compatibility conditions~\eqref{eq:mo-1}-\eqref{eq:mo-3} of \eqref{eq:3-term-r} hold.
\end{prop}
\begin{rem}
	The first equation \eqref{eq:dot-a-r} follows directly from \eqref{eq:dot-D-r}, and the second equation \eqref{eq:dot-b-r} is a consequence of \eqref{eq:D-mT}.
\end{rem}

\section{Conclusion}
To conclude the discussion on mutual dependencies between Hermite--Pad\'{e} approximants, integrability, and multiple orthogonal polynomials let us pre\-sent two directions of further studies. Each of them starts from one known extension of a particular ingredient of the three given, what motivates to find analogous generalizations within the two other subjects.

\subsection{Non-commutative integrable discrete systems and Hermite--Pad\'{e} approximation} 
Searching for a non-commutative extension of a given integrable discrete system is well motivated by physics. For the Hirota equations such a genera\-lization was given in \cite{FWN-Capel,Nimmo-NCKP}, see also~\cite{DoliwaKashaev} for its application in mathematical physics. The basic technical tool there is provided by the notion of quasideterminant~\cite{Quasideterminants-GR1,Quasideterminants}. The idea to transfer the connection between Pad\'{e} approximation, orthogonal polynomials and Toda equations is present in~\cite{Quasideterminants}, but see also earlier works~\cite{Draux-OP-PA,Draux-rev} and further developments \cite{Doliwa-NCCF,Doliwa-Siemaszko-1} motivated by both geometric approach to integrability and theory of formal languages. 

In \cite{Doliwa-NHP} we formulated the Hermite--Pad\'{e} problem for series with non-commuting coefficients and found its solution in terms of quasi-determinants. The resulting equations, which include the non-commutative version of the Paszkowski constraint~\eqref{eq:Paszkowski-Frobenius-DD}, motivated us to propose integrable non-commu\-ta\-ti\-ve generalization of the multidimensional discrete Toda equations. 

\subsection{Generalization from approximation to interpolation problems}
Already Mahler \cite{Mahler-P} studied the interpolation analogs of both the Hermite--Pad\'{e} and simultaneous Pad\'{e} approximation problems and gave the corresponding version of Theorem~\ref{th:Mahler}. In the confluent case when all the interpolation nodes coincide, and with transition to the appropriate tangency condition, one obtains the approximation problems. In \cite{Doliwa-NAMTS} we solved the Hermite--Pad\'{e} interpolation problem in terms of a generalization of determinantal formulas \eqref{eq:Z-det}. In the simplest case one obtains this way classical Cauchy solution~\cite{Cauchy,CuytWuytack} of the  rational interpolation problem. 

Our solution satisfies non-autonomous versions, where the interpolation nodes appear explicitly, of the corresponding equations of the Hermite--Pad\'{e} approximation theory. In particular, the interpolation generalization of the Paszkowski constraint gives rise to integrable non-autonomous multidimensional Toda equations. In the simplest case of dimension two the equations reduce to the known non-autonomous discrete-time Toda equations~\cite{SpiridonovZhedanov,Hirota-naTl,KajiwaraMukaihira}.

\subsection*{Acknowledgment}
I would like to thank the organizers of the XL WGMP conference for the invitation and support.

% ------------------------------------------------------------------------

\begin{thebibliography}{spmpsci}
\bibliographystyle{cite}

\bibitem{AblSi}
Ablowitz, M.J., Segur, H.: Solitons and the Inverse Scattering, SIAM
Philadelphia (1981)


\bibitem{AdlervanMoerbeke}
Adler, M., van Moerbeke, P.: Generalized orthogonal polynomials, discrete KP and Riemann-Hilbert problems. Comm. Math. Phys. \textbf{207},  589--620 (1999)

\bibitem{AdlervanMoerbekeVanhaecke}
Adler, M., van Moerbeke, P., Vanhaecke, P.: Moment matrices and multi-component KP,
	with applications to random matrix theory.
Commun. Math. Phys. \textbf{286},  1--38 (2009)

\bibitem{Akhiezer}
Akhiezer, N.I.: The classical moment problem and some related questions in analysis. Olivier \& Boyd, Edinburgh and London (1963)

\bibitem{Aptekarev}
Aptekarev, A. I.: Multiple orthogonal polynomials. J. Comput. Appl. Math. \textbf{99},  423--447 (1998)


\bibitem{AptekarevDerevyaginVanAssche}
Aptekarev, A.I., Derevyagin, M., Van Assche, W.: Discrete integrable systems generated by Hermite--Pad\'{e} approximants. Nonlinearity \textbf{29},  1487--1506 (2016)	


\bibitem{AptekarevDerevyaginMikiVanAssche}
Aptekarev, A.I., Derevyagin, M., Miki, H., Van Assche, W.: Multidimensional Toda lattices: continuous and discrete time. SIGMA \textbf{12}, 054 (2016)

\bibitem{Baker}
Baker, Jr., G.A., Graves-Morris, P.: Pad\'{e} approximants. Cambridge University Press, Cambridge (1996)

\bibitem{BeckermannLabahn}
Beckermann, B., Labahn, G.: Fraction-free computation of simultaneous Pad\'{e} approximants. In: Proceedings of the 2009 International Symposium on Symbolic and Algebraic Computation ISSAC'09, pp. 15--22. ACM, New York (2009)

\bibitem{BleherKuijlaars}
Bleher, P.M., Kuijlaars, A.B.J.: Random matrices with external source and multiply orthogonal polynomials. Int. Math. Res. Not. \textbf{2004}, 109--129 (2004)


\bibitem{Brezinski}
Brezinski, C.: History of continued fractions and Pad\'{e} approximants. Springer, Berlin (1991)

\bibitem{Brezinski-PTA-OP}
Brezinski, C.: Pad\'{e}-type approximation and general orthogonal polynomials. Springer, Basel AG (1980)


\bibitem{BultheelvanBarel}
Bultheel, A., van Barel, M.: Linear algebra, rational approximation, and orthogonal polynomials. Studies in computational mathematics Vol. 6, Elsevier (1997)

\bibitem{Cauchy}
Cauchy, A.L.: Cours d`Analyse de l`Ecole Royale Polytechnique. Premi\`{e}re Partie. Analyse alg\'{e}braique. Imprim\'{e}rie Royale, Paris (1821)


\bibitem{Chihara}
Chihara, T.S.: An Introduction to Orthogonal Polynomials. Gordon and Breach, New York (1978)

\bibitem{CuytWuytack}
Cuyt, A., Wuytack, L.: Nonlinear methods in numerical analysis. North Holland Publishing Company, Amsterdam - New York - Oxford - Tokyo (1987)

\bibitem{Darboux-TS}
Darboux, G.: Le\c{c}ons sur la th\'{e}orie g\'{e}n\'{e}rale des surfaces. {I--IV}. Gauthier -- Villars, Paris (1887--1896)


\bibitem{DKJM}
Date, E., Kashiwara, M., Jimbo, M., Miwa, T.: Transformation groups for
	soliton equations. In: Jimbo, M.,  T. Miwa, T.  (eds.) Nonlinear integrable systems --- classical theory and
quantum theory, Proc. of RIMS Symposium, pp. 39--119. World
Scientific, Singapore (1983)

\bibitem{DJM-II}
Date, E., Jimbo, M., Miwa, T.: Method for generating discrete soliton
	equations. II. J. Phys. Soc. Japan \textbf{51}, 4125--4131 (1982) 
	
\bibitem{Deift}
Deift, P.A.: Orthogonal Polynomials and Random Matrices: A Riemann-Hilbert Approach. AMS, New York (1998)


\bibitem{Iglesia}
de la Iglesia, M.D.: Orthogonal Polynomials in the Spectral Analysis of Markov Processes. Cambridge University Press, Cambridge (2022)	

\bibitem{DD-DC-1}
Della Dora, J., Di Crescenzo, C.: Approximants de Pad\'{e}--Hermite. 1\`{e}re partie: theorie. Numer. Math. \textbf{43},  23--39 (1984), 

\bibitem{DD-DC-2}
Della Dora, J., Di Crescenzo, C.: Approximants de Pad\'{e}--Hermite. 2\`{e}me partie: programmation. Numer. Math. \textbf{43}, 41--57 (1984)

\bibitem{DCN}
Doliwa, A.: Geometric discretisation of the Toda system.
Phys. Lett. A \textbf{234}, 187--192 (1997)

\bibitem{Dol-Des} 
Doliwa, A.: Desargues maps and the Hirota--Miwa equation. Proc. R. Soc. A
\textbf{466},  1177--1200 (2010)

\bibitem{Dol-AN} 
Doliwa, A.: The affine Weyl group symmetry of Desargues maps and of the non-commutative Hirota--Miwa system. Phys. Lett. A {\bf 375},  1219--1224 (2011)



\bibitem{Doliwa-NCCF}
Doliwa, A.: Non-commutative double-sided continued fractions. J. Phys. A: Math. Theor. \textbf{53}, 364001 (2020)


\bibitem{Doliwa-NHP}
Doliwa, A.: Non-commutative Hermite--Pad\'{e} approximation and integrability. Lett. Math. Phys. \textbf{112}, 68 (2022)

\bibitem{Doliwa-NAMTS}
Doliwa, A.: Non-autonomous multidimensional Toda system and multiple interpolation problem. J. Phys. A: Math. Theor. \textbf{55}, 505202  (2022) 


\bibitem{DoliwaKashaev}
Doliwa, A., Kashaev,  R.M.: Non-commutative bi-rational maps satisfying Zamolodchikov equation, and Desargues lattices. J. Math. Phys. \textbf{61},  092704 (2020)


\bibitem{Doliwa-Siemaszko-1}
Doliwa, A., Siemaszko, A.: Integrability and geometry of the Wynn recurrence. Numer. Algorithms \textbf{92}, 571--596 (2023) 

\bibitem{Doliwa-Siemaszko-2}
Doliwa, A., Siemaszko, A.: Hermite--Pad\'{e} approximation and integrability. J. Approx. Theory \textbf{292}, 105910 (2023)


\bibitem{Draux-rev}
Draux, A.: The Pad\'{e} approximants in a non-commutative algebra and their applications. In: Werner, H.,  B\"{u}nger, H. J. (eds) Pad\'{e} Approximation and its Applications, Bad Honnef 1983. Lecture Notes in Mathematics \textbf{1071} Springer, Berlin, Heidelberg (1984) 

\bibitem{Draux-OP-PA}
Draux, A.: Formal orthogonal polynomials and Pade approximants in a non-commutative algebra. In: Fuhrmann, P.A. (ed.) Mathematical Theory of Networks and Systems, Lecture Notes in Control and Information Sciences \textbf{58}. Springer, Berlin, Heidelberg (1984)

\bibitem{Flaschka}
Flaschka, H.: The Toda lattice. I.
Existence of integrals, Phys. Rev. B  \textbf{9}, 1924--1925 (1974)

\bibitem{Frobenius}
Frobenius, G.: Ueber Relationen zwischen den Näherungsbrüchen von Potenzreihen. J. reine und angew. Math. (Crelle's Journal) \textbf{90}, 1--17 (1881)


\bibitem{GGKM}
Gardner, C.S., Greene, J.M., Kruskal, M.D., Miura, R.M.:
Method for Solving the Korteweg--deVries Equation.
Phys. Rev. Lett. \textbf{19}, 1095--1097 (1967) 


\bibitem{Quasideterminants-GR1}
Gelfand, I., Retakh, V.: Theory of noncommutative determinants
	and characteristic functions of graphs. Funct. Anal. Appl. \textbf{26},  1--20 (1992)

\bibitem{Quasideterminants}
Gelfand, I., Gelfand, S., Retakh, V., Wilson, R.L.: Quasideterminants. Adv. Math. \textbf{193}, 56--141 (2005) 

\bibitem{Geronimus}
Geronimus,	Ya.: Polynomials orthogonal on a circle and interval. Pergamon Press, Oxford (1960)



\bibitem{Gragg}
Gragg, W.B.: The Pad\'{e} table and its relation to certain algorithms of numerical analysis. SIAM Review \textbf{14}, 1--62 (1972)

\bibitem{GRP}
Grammaticos, B., Ramani, A., Papageorgiou, V. G.: Do integrable mappings have the Painlev\'{e} property? Phys. Rev. Lett. \textbf{67},  1825--1828 (1991)


\bibitem{Hermite}
Hermite, C.: Sur la fonction exponentielle. Oeuveres~III, 150--181 (1873) 

\bibitem{Hermite-P}
Hermite, C.: Sur la g\'{e}n\'{e}ralisation des fractions continues alg\'{e}briques. Oeuveres~IV, 357--377 (1893)

\bibitem{IDS}
Hietarinta, J., Joshi, N., Nijhoff, F. W.: Discrete systems and integrability. Cambridge University Press, Cambridge (2016)

\bibitem{Hirota-2dT}
Hirota, R.: Nonlinear partial difference equations. II. Discrete-time Toda equation. J. Phys. Soc. Japan, \textbf{43},  2074--2078 (1977)

\bibitem{Hirota}
Hirota, R.: Discrete analogue of a generalized Toda equation. J. Phys. Soc. Jpn. \textbf{50}, 3785--3791 (1981)


\bibitem{Hirota-naTl}
Hirota, R.: Conserved quantities of ``random-time Toda equation''. J. Phys. Soc. Japan \textbf{66},  283--284 (1997)

\bibitem{Hirota-book}
Hirota, R.: The direct method in soliton theory.
Cambridge University Press, Cambridge (2004)

\bibitem{Hirota-1993}
Hirota, R., Tsujimoto, S., Imai, T.: Difference scheme of soliton equations. In: Christiansen, P.L., Eilbeck, P.L., Parmentier, R.D. (eds.) Future Directions of Nonlinear Dynamics in Physical and Biological Systems, pp.~7--15. Springer (1993)

\bibitem{Ismail}
Ismail, M.E.H.: Classical and Quantum Orthogonal Polynomials in One Variable.  Cambridge University Press, Cambridge (2005)

\bibitem{Jacobi}
Jacobi, C. G.: \"{U}ber die Darstellung einer Reihe gegebner Werthe durch eine gebrochene rationale function. J. Reine Angew. Math. \textbf{30},  127--156 (1846)

\bibitem{Kac-vMoerbeke}
Kac, M., van Moerbeke, P.: On an explicitly soluble system of nonlinear differential equations related to certain Toda lattices. Adv. Math. \textbf{16}, 160--169 (1975)

\bibitem{KajiwaraMukaihira}
Kajiwara, K., Mukaihira, A.: Soliton solutions for the non-autonomous discrete-time Toda lattice equation. J.~Phys.~A: Math. Gen. \textbf{38}, 8727--8737 (2005)

\bibitem{KarlinMcGregor}
Karlin, S., McGregor, J.L.: The differential equations of birth-and-death processes, and the
Stieltjes moment problem. Trans. Amer. Math. Soc. 
 \textbf{85}, 489--546 (1957)

\bibitem{Klimyk-Schmudgen}
Klimyk, A., Schm\"{u}dgen, K.: Quantum Groups and Their Representations. Springer, Berlin (1997)

\bibitem{Kuijlaars}
Kuijlaars, A.B.J.: Multiple orthogonal polynomial ensembles. Recent trends in orthogonal polynomials and approximation theory. Contemp. Math. \textbf{507}, 155--176 (2010) 


\bibitem{KNS-rev}
Kuniba, A., Nakanishi, T., Suzuki, J.: $T$-systems and $Y$-systems in integrable systems. 
J. Phys. A: Math. Theor. {\bf 44}, 103001 (2011) 

\bibitem{LedermannReuter}
Ledermann, W., Reuter, G.E.H.: Spectral theory for the differential equations of simple
birth and death processes. Philos. Trans. Roy. Soc. London A \textbf{246}, 321--369 (1954)


\bibitem{Mahler}
Mahler, K.: Zur Approximation der Exponentialfunktion und des Logarithmus, Teil I. J. Reine Angew. Math. \textbf{166},  118--150 (1932)

\bibitem{Mahler-P}
Mahler, K.: Perfect systems. Compositio Math. \textbf{19},  95--166 (1968)

\bibitem{ManoTsuda}
Mano, T., Tsuda, T.: Hermite--Pad\'{e} approximation, isomonodromic deformation and hypergeometric integral. Math. Zeitschrift \textbf{285},  397--431 (2016)

\bibitem{Mikhailov}
Mikhailov, A.V.: Integrability of the two-dimensional generalization of Toda chain. JETP Lett. \textbf{30} 414--418 (1979)

\bibitem{Miwa} 
Miwa, T.: On Hirota's difference equations. 
Proc. Japan Acad. \textbf{58}, 9--12 (1982) 

\bibitem{Moser}
Moser, J.: Three integrable Hamiltonian systems connected with isospectral deformations. Adv. Math. \textbf{16}, 197--220 (1975) (1975)

\bibitem{NagaiTokihiroSatsuma}
Nagai, A., Tokihiro, T., Satsuma, J.: The Toda molecule equation and the $\varepsilon$-algorithm. Mathematics of Computation, \textbf{67}, 1565--1575 (1998)


\bibitem{Nagao-Yamada}
Nagao, H., Yamada, Y.: Pad\'{e} Methods for Painlev\'{e} Equations. Springer, Singapore (2021)

\bibitem{FWN-Capel}
Nijhoff, F.W., Capel, H.W.: The direct linearization approach to
	hierarchies of integrable PDEs in $2+1$ dimensions: I. Lattice equations and the
	differential-difference hierarchies. Inverse Problems \textbf{6}, 567--590 (1990)


\bibitem{Nikiforov-Suslov-Uvarov}
Nikiforov, A.F., Suslov,  S.K., Uvarov,  V.B.: Classical Orthogonal Polynomials of a Discrete Variable.
Springer-Verlag, Berlin-Heidelberg (1991)


\bibitem{NikishinSorokin}
Nikishin, E.M.,  Sorokin, V.N.: Rational approximation and orthogonality. Transl. Math. Monographs \textbf{92}, Amer. Math. Soc. (1991)



\bibitem{Nimmo-NCKP}
Nimmo, J.J.C.: On a non-Abelian Hirota-Miwa equation.
J. Phys. A: Math. Gen. \textbf{39}, 5053--5065 (2006) 

\bibitem{Olshanetsky-Perelomov}
Olshanetsky, M.A.,Perelomov,  A.M.: The Toda chain
as a reduced system. Teor. Mat. Fis. \textbf{45}, 3--18 (1980)

\bibitem{Olver}
Olver, P.: Application of Lie Groups to Differential Equations, Springer, Berlin -- New York (1986)


\bibitem{Pade}
Pad\'{e}, H.: Sur la r\'{e}presentation approch\'{e}e d'une fonction par des fractions rationelles, Ann. \'{E}cole Nor. \textbf{9}, 1--93 (1892)

\bibitem{PGR-LMP}
Papageorgiou, V., Grammaticos, B., Ramani, A.: Orthogonal polynomial approach to discrete Lax pairs for initial boundary-value problems of the QD algorithm. Lett. Math. Phys. \textbf{34}, 91--101 (1995)

\bibitem{Paszkowski}
Paszkowski, S.: Recurrence relations in Pad\'{e}--Hermite approximation. J. Comput.
Appl. Math. \textbf{19}, 99--107  (1987) 

\bibitem{Schoutens}	
Schoutens, W.: Stochastic Processes and Orthogonal Polynomials. Springer, New York (2000)


\bibitem{Shiota}
Shiota, T.: Characterization of Jacobian varieties in terms of soliton equations. Invent. Math. \textbf{83}, 333--382  (1986)


\bibitem{SpiridonovZhedanov}
Spiridonov, V., Zhedanov, A.: Discrete Darboux transformations, the discrete-time Toda lattice, and the Askey--Wilson polynomials. Math. Appl. Anal. \textbf{2}, 369--398 (1995) 


\bibitem{Symes}
Symes, W.W.: The QR algorithm and scattering for the finite nonperiodic Toda lattice. Physica D \textbf{4}, 275--280 (1982)


\bibitem{Szego}
Szeg\H{o}, G.: Orthogonal polynomials. AMS, Providence, RI, fourth edition (1975)


\bibitem{Toda-TL}
Toda, M.: Waves in nonlinear lattice.
Progr. Theoret. Phys. Suppl. \textbf{45}, 174--200 (1970)

\bibitem{VanAsche-HPAO}
Van Assche, W.: Pad\'{e} and Hermite--Pad\'{e} approximation and orthogonality. Surv. Approx. Theory \textbf{2}, 61--91 (2006)

\bibitem{VanAsche-nn-mop}
Van Assche, W.: Nearest neighbor recurrence relations for multiple orthogonal polynomials. J. Approx. Theory \textbf{163}, 1427--1448 (2011)

\bibitem{VanAssche}
Van Assche, W.: Orthogonal polynomials and Painlev\'{e} equations. Cambridge University Press (2018)


\bibitem{Moerbeke}
Van Moerbeke, P.: The Spectrum of Jacobi Matrices. Inventiones Math. \textbf{37}, 45--81 (1976) 

\bibitem{Vilenkin-Klimyk}
Vilenkin, N.Ja., Klimyk,  A.U.: Representation of Lie Groups and Special Functions: Volume 1: Simplest Lie Groups, Special Functions and Integral Transforms, Volume 2: Class I Representations, Special Functions, and Integral Transforms, Volume 3: Classical and Quantum Groups and Special Functions. Kluwer Academic Publishers (1991, 1993, 1992)


\bibitem{ZMNP}
Zakharov, V.E., Manakov, S.V., Novikov, S.P.,  Pitaievsky, L.P.: Theory of
	Solitons: the Inverse Scattering Method, Plenum, New York (1984) 


\end{thebibliography}
\end{document}